\documentstyle[prb,aps,psfig,multicol]{revtex}
\begin{document}
\title{\bf{ DYNAMICS OF UNBINDING OF POLYMERS IN A RANDOM MEDIUM} }
\author{Somendra M. Bhattacharjee}
\address{Institute of Physics, Bhubaneswar 751 005, India}
\author{A. Baumg\"artner}
\address{Forum Modellierung, Forschungzentrum, J\"ulich D51245 Germany}
\maketitle
\widetext
\begin{abstract}
We have studied the aging effect on the dynamics of unbinding of a
double stranded directed polymer in a random medium. By using the
Monte Carlo dynamics of a lattice model in two dimensions, for which
disorder is known to be relevant, the unbinding dynamics is studied by
allowing the bound polymer to relax in the random medium for a waiting
time and then allowing the two strands to unbind.  The subsequent
dynamics is formulated in terms of the overlap of the two strands and
also the overlap of each polymer with the configuration at the start
of the unbinding process.  The interrelations between the two and the
nature of the dependence on the waiting time are studied.
\end{abstract}
\begin{multicols}{2}
\section{Introduction}

The dynamics of polymers near a phase transition point, especially in
presence of randomness or disorder, is important in many situations
like denaturation of DNA, protein folding, collapse of heteropolymers
etc \cite{physrept,pf}.  In addition, polymers and extended elastic
manifolds in random media constitute a class of problems which appear
in various disguise in many problems, as e.g., interfaces in random
systems, flux lines in high Tc superconductors, surface growths and
others\cite{halp}.  Certain aspects of dynamical properties for
polymeric objects have been discussed in the past, with emphasis
mostly on the equilibrium or stationary dynamics\cite{ab96}.  In
random systems, off-equilibrium dynamics has a special role because
the system has to explore the phase space in search of its equilibrium
state, if it reaches there at all\cite{fischer}.  Thus, the
off-equilibrium dynamics near a phase transition is expected to be
different from the pure dynamics. In this paper, we study a very
simple polymer model with a phase transition for which equilibrium
properties are known with a certain degree of confidence.  The
particular model we study is the unbinding transition of two
interacting directed polymers in a random medium.  This corresponds to
a simplified model of denaturation of DNA in a solution with quenched
random impurities\cite{physrept,peyard}.

Even though the off-equilibrium dynamics in glassy polymers are known
for a long time\cite{struik}, the peculiarities of dynamics of random
systems received attention rather recently through experiments on
various systems \cite{ll,angell}.  As yet, there is no analytical
approach for these problems, but several conflicting scenarios have
been suggested, with the lack of well accepted equilibrium theories
adding to the sore. In this respect, the model we are considering is
in a rather enviable position, because of several analytical tools and
results available for equilibrium properties.

A $(d+1)$ dimensional directed polymer (DP) is a polymer with a
preferred direction so that it has random fluctuation in the
transverse $d$ directions only.  Such an interacting DP system with
homogeneous interaction has been proposed in the past for denaturation
of DNA \cite{peyard} in a pure solvent, where the most important
degree of freedom taken into account is the interstrand base pairing.
Our model includes a quenched distribution of impurities in the
environment \cite{rani}.

The main effect of randomness in dynamics is the aging
effect\cite{fischer,slow}.  If the system is allowed to equilibrate
upto a certain time $t_w$, to be called the waiting time, then the
subsequent dynamics under a perturbation depends on this imposed time
$t_w$ in a nontrivial way.  We like to explore this aging effect in
the dynamics of unbinding through a Monte Carlo dynamics of a lattice
model.

We discuss below in section 2 the equilibrium properties of this
interacting system of two polymers.  There we also point out the
connection of this two chain problem with that of a single chain via
the replica approach.  We then discuss the methodology of our
simulation in section 3. The results are presented and discussed in
section 4.

\section{Equilibrium behavior}
Let us consider two DP in the same random medium so that the
Hamiltonian in a path integral approach can be written as\cite{sm1}
\begin{eqnarray}
&&{\cal H}=\sum_{i=1}^{i=2}
\int_0^N \ dz\ [\frac{1}{2}{\dot{\bf r}}_i(z)^2 +
V({\bf r}_{i}(z),z) ]
\ +\nonumber\\
&&\ \ \  \int_0^N dz\ v\
\delta({\bf r}_{1}(z) -  {\bf r}_{2}(z)) ,
\label{eq:ham} 
\end{eqnarray}
where ${\bf r}_i(z)$ denotes the $d$ dimensional transverse spatial
coordinate of the $i$th polymer at contour length $z$, and ${\dot{\bf
r}}_i(z) = {\partial {\bf r}_i(z)}/{\partial z}$. The first term
denotes the elastic energy part of the Gaussian chains and the second
term is the random potential $V({\bf r},z) $ at point $({\bf r},z)$,
and the last term denotes the mutual contact interaction between the
chains. Note that the interaction is always at equal length
\cite{peyard}.

It is known that randomness is relevant in $d=1$ dimension
\cite{kardar,halp} and the polymer has to swell to take advantage of
the favorable energy pockets.  The transverse size grows with the
length with an exponent $\nu = 2/3$ which is bigger than the Gaussian
value $1/2$ (expected for the pure case even in presence of the
interaction.

This particular problem of two interacting chain in the same random
medium was considered numerically by Mezard in an attempt to calculate
the overlap of two replicas for the single chain problem, the overlap
being the most important quantity in a replica approach\cite{mezard}.
A general formulation for any $d$ was given by Mukherji who, in
addition to establishing the exact exponent for overlap in the 1+1
case, also obtained the relevant exponents for $d=2+\epsilon$ for the
spin glass transition point.  This formulation was also used to study
higher order overlaps\cite{sm2}, and in the strong coupling
phase\cite{kinzel} for $d>2$.

In a dynamic renormalization group approach, Mukherji\cite{sm1} showed
that the interaction is relevant in all dimensions.  Each chain
individually behaves as in the single chain problem, i.e. the relevant
strong disorder fixed point is independent of $v$.  A straight forward
extension of the approach of ref\cite{sm1} gives the nontrivial fixed
point for the two repelling (i.e., unbound) chains in the random
medium\cite{kinzel}.  The fixed point diagram is shown in fig 1, that
shows that $v=0$ remains the critical point for the binding-unbinding
transition for the two chains. A bound state forms for $v<0$.  The
relevant exponents are also obtainable from the RG recursion
relations.

The order parameter that describes the critical point is the overlap
or the number of contacts of the two chains, defined as
\begin{equation}
q(v) = \frac{1}{N} \ \ \int_0^N dz\ v\
\delta({\bf r}_{1}(z) -  {\bf r}_{2}(z)).
\label{eq:qv}
\end{equation}
The scaling behavior found for this overlap is $q = f(v N^{2/3})$ for
polymers of length $N$ near the $v^*=0$ fixed point\cite{sm1}.  This
particular scaling can be justified by a simple argument.  An overlap
on a length scale $L$ along the chain costs an energy $v L$ while the
gain from free energy fluctuation by following two different paths on
this length $L$ is $L^{\chi}$, with $\chi=1/3$ for this 1+1
dimensional problem.  This gives the scaling variable $vL^{2/3}$ as
obtained exactly in Ref. \cite{sm1}. We generalize this argument below
for dynamics.  The excitations we are considering here are the loops
on a scale $L$ and this forms the basis of the droplet picture for
DP\cite{hff}.  For large $N$, the argument approaches the nontrivial
fixed point, and being the unbound phase, $q=0$, with a finite size
scaling form $q=f(\Delta v N^{-2/3})$.  $\Delta v$ is the deviation
from the fixed point.

If we consider the single chain problem, then the overlap, in the
replica approach, is given by this $q$ at $v=0$.  Though this quantity
is not available from RG, numerical computations\cite{mezard,bar} show
that $q\neq 0$.  This gives the Edwards-Anderson order parameter for
this strong disorder phase (see below).  We therefore see that the
order parameter for the critical point is a simple generalization of
the order parameter needed for a replica approach of the single chain
problem.  In this respect, this DP problem is unique among the known
random models.
 
In spite of these results for the equilibrium behavior, very little is
known about the dynamics of unbinding, though certain aspects of the
single chain dynamics have recently been looked into\cite{yoshi,bar}.
Our aim is to explore the time evolution of the overlap for the
unbinding transition, and the effect of aging on this evolution, and
correlate with the single chain behavior.

\section{Model and Method}

To study the dynamics, we consider DPs on a square lattice.  The
polymers start at the origin and are allowed to take steps only in the
$+x$ or $+y$ directions without any a priori bias.  This produces
polymers directed along the diagonal of the lattice.  Two polymers
interact if they share a point and each contact is assigned an energy
$u$.  In addition, there is a random energy at each site chosen from a
uniform distribution ${\bar\epsilon} \in $[-.5,.5].  At a given
temperature, there are two parameters,$v=u/k_BT$ and $\epsilon =
{\bar\epsilon}/k_BT$.  We use the standard Metropolis single bead flip
for the Monte Carlo dynamics\cite{baumg1}. The chains are always
anchored at one end but free at the other\cite{comm}.  At each step
the bead to be moved is chosen randomly from the $2 N -2 $ beads. One
MC time step then consists of $2N -2 $ such attempts.  The dynamics is
performed for a given disorder realization, averaged over several
random number realizations (thermal average)and initial configuration,
and then averaged over disorder realizations.

Our procedure involves two chains completely bound (on top of each
other) together evolving in the random potential for a time $t_w$
(i.e. MC is done with respect to random energy only) and then the
chains evolve individually in presence of the interaction also. See
Fig 1b.  With respect to the fixed point diagram of Fig 1a, the bound
double stranded chain evolves towards the ``strong disorder'' fixed
point $K$ upto time $t_w$, and after that the evolution is towards the
stable fixed point $M$.  We monitor the average fraction of contacts
(overlaps) of the two chains and the overlap of each chain with the
configuration at time $t_w$.

Let us define two quantities self overlap $C_i$ and mutual overlap $q$
as
\begin{mathletters}
\begin{eqnarray}
&&C_i(v,t+t_w)=\frac{1}{N} \sum {\overline{\langle \delta({\bf r}_i(t+t_w) - {\bf
r}_i(t_w))\rangle }},\nonumber\\
&&q(v,t+t_w) = \frac{1}{N} \sum {\overline{\langle
\delta({\bf r}_1(t+t_w) - {\bf r}_2(t+t_w))\rangle }},
\end{eqnarray}
\end{mathletters}
where $\langle..\rangle$ denotes thermal average and overbar denotes
disorder average.  The mutual overlap $q(v,t)$ defined here is a time
dependent generalization of the equilibrium overlap of
Eq. \ref{eq:qv}, while $C_i$ is the overlap of the configuration of
chain $i$ at time $t+t_w$ with its configuration at time $t_w$.  By
symmetry $C_i$ is independent of the chain index $i$.

It is also possible to relate the overlap $q$ to a correaltion
function.  Let us define $s_i(t)=1$ if at time $t$ there is an overlap
of the two chains at chain length $i$, otherwise it is zero.  The
overlap at time $t$ is then $\sum_i s_i(t)/N$.  If we define an
autocorrelation function ${\cal C}(t_1,t_2) = N^{-1} \sum_i
{\overline{\langle s_i(t_1)s_i(t_2)\rangle}} $, we see that
$q(v,t+t_w) = {\cal C}(t_w,t+t_w)$ because $s_i(t_w)=1$ forall $i$.

For the single chain problem the self-overlap,$C_i$, defined above is
also a quantity of fundamental importance.  If we take limit $t_w
\rightarrow \infty$ first and then $t\rightarrow \infty$, then for
$v=0$, $C_i$ would correspond to the Edwards-Anderson order parameter
for the strong disorder phase.  This is because $C_i$ would then
measure the overlap, in equilibrium, of the polymer configuration at
two widely spaced time, and a nonzero value would imply a frozen
random configuration, characteristic of a ``strong disorder'' phase.
We therefore expect
\begin{equation}
\lim_{t\rightarrow\infty}\ \lim_{t_w\rightarrow\infty} C(0,t+t_w) =
q_{EA}.
\end{equation}
In fact, for $v=0$, in the limit $t_w\rightarrow \infty$, one can also
connect this overlap with the the self overlap defined above as
$q(0,t) = C(0,2 t)$ because in the equilibrium, the overlap of the two
configurations for $C$ will be the same as the overlap needed for $q$.
This is a check on our simulation for $t\ll t_w$.

In the simulation, $q$ and $C$ were monitored for various values of
$v$, and $t_w$, for chains of length upto 300.  At this length the
dynamics we report here do not have significant finite size
effects. Note also that by construction there is no finite size effect
in the transverse direction.

\section{Results and discussions}

We show the results of the simulation in Fig 2, where the overlap for
various waiting times and $v$ are plotted.  Fig 2A shows the results
for the pure system ($\bar\epsilon =0$) for $v=1.0$, and there is no
significant dependence on the waiting time. For the random case, shown
in Fig 2B, we see the longer the waiting time the slower the
relaxation. In other words, the system develops a stiffness as it ages
in the random environment.  This is the first effect of ``aging''.  In
absence of detailed theories, we considered various scaling forms. The
form used for the single chain problem in Ref. \cite{yoshi} turns out
to be applicable in this interacting problem.  A data collapse is
obtained by plotting $c q(v,t+t_w)$ vs. $t/t_w$ with suitable choices
of the prefactor $c$.  The variation of $c$ with $t_w$ and $v$ is
shown in the inset in Fig 2C. Similar scaling is obtained for the self
overlap also (not shown). However, there seems to be no
``universality'' in the sense that the dynamics do depend upon the
strength of interaction.  It is not possible to go to large values of
repulsion in this 1+1 dimensional problem because of the log-jamming
problem on a lattice.

For $t\ll t_w$, the early time dynamics is the ``quasi''-equilibrium
dynamics. For the largest $t_w$,we find a linear relationship with the
self overlap and the slope decreases with increasing $v$.  Fig 3A
shows the early time equality, $q(0,t)=C(0,2t)$ for $v=0$ and its
failure for $v \neq 0$.  In fact, if we assume that for early times,
$t\ll t_w$, $C(t) \approx t^{-x}$, then, one can write $q(0,t) \equiv
C(2t) = 2^{-x} C(t)$.  Assuming such a homogeneity relation for
nonzero $v$, we can write $q(v,t) = b C(v,t)$ so that by choosing the
coefficient, $b \ (= 2^{-x}\ {\rm for} \ v=0)$ it would be possible to
get a data collapse for all $v$ at least for early times.  We do see
such a collapse at early times as shown in Fig 3.  This indicates a
power law behavior, and we conclude that the early time power law
decay of the overlap has the same exponent as the self overlap.

Combining the various forms, a scaling formula for aging can be
suggested as
\begin{equation}
q(v,t+t_w) = t^{-x} f(t/t_w)
\end{equation}
which for the limit $t_w\rightarrow \infty$, and then $t
\rightarrow\infty$, would give $q=0$ and not a finite $q_{EA}$.  Such
a form has been used for various random systems in spite of this
problem\cite{slow}, and numerical simulations are yet to sort this
out\cite{reiger}. Fig 3 suggests that a similar equation is valid for
the self-overlap \cite{yoshi} with the same, rather small, exponent
$x$.

For the largest $t_w$, we see a power law decay of the overlap at
early times and not a logarithmic decay as would be expected from the
droplet picture\cite{hff}.  In the droplet picture one assumes that
the dynamics is governed by the typical barrier, and hence is of
activated type.  So, on a time scale $t$, the system would explore the
phase space on length scales for which the barrier heights $B \sim \ln
t$.  If one assumes further a growth of barriers with length scale
$B\sim L^{\psi}$, then the relevant length scale at time $t$ is
$L(t)\sim (\ln t)^{1/\psi}$.  If we now generalize the scaling picture
mentioned in section 2 to dynamics with the hypothesis that the
dynamics is governed by the length scale $L(t)$ at that time, one
would expect a dynamic scaling
\begin{equation}
q(v,t) = q(v L(t)^{2/3})= q( v (\ln t)^{(1-\chi)/\psi})
\end{equation}
This is valid for $\ln t \ll \ln t_w$.  The simulation results are
then not consistent with this dynamic scaling.  In fact, no MC
simulations have so far produced this log time scale in early dynamics
in random systems.  It has been speculated that the power law form,
instead of logarithm of time, is a consequence of a logarithmic growth
of barrier heights as opposed to $B\sim L^{\psi}$.  However, this is
ruled out for DP, because it is known from transfer matrix
calculations \cite{mikh} that, in the 1+1 dimensional case, the
typical barrier has the same scaling form as the free energy
fluctuation, $\psi=\chi=1/3$ .  It is possible that the early dynamics
is not controlled by the typical barriers but rather by the smallest
barriers.  If we denote the probability distribution of barrier
heights by $P(B)$, and if $P(B)$ diverges (but integrable) as
$B\rightarrow 0$ then early dynamics would not be activated type but
rather like in spinodal decomposition where barrierless diffusion is
the relevant mechanism.  In such situations one finds that the time
scale is a power law in the barrier height\cite{shenoy} as observed in
simulations here.
 
In terms of lengthscales, the combined (bound) chains equilibrate by
crossing barriers over lengthscales $L(t_w)$, length being measured
along the chain.  The subsequent unbinding then involves the
separation of the chains in presence of the repulsion within this
length scale, $L(t) \ll L(t_w)$.  Once $t \approx t_w$, one observes
true nonequilibrium decay.  Our data suggest again a power law but the
overall decay of the overlap is rather small to get a reliable
estimate of the exponent or any other functional form.  However a
scaling variable $L(t)/L(t_w)$ seems to be a natural choice, which we
find to be related to $t/t_w$.  This also indicates that the relation
between $L(t)$ and $t$ should be a power law type.  We would like to
add that there is the possibility that the length and time scales
studied in these lattice simulations of this paper may not be in the
right asymptotic limit to observe the dynamics predicted by the
droplet picture.  In fact, more analytical work is necessary to
understand the finite size and crossover effects in early dynamics of
random systems in general.

A bound on the late time decay of the overlap can be obtained by
considering each bead independently (i.e. not connected as a polymer).
In this case the overlap $q$ is just the probability of reunion of two
vicious walkers (repulsive random walkers) at time
$t$\cite{fisher,reun}.  This probability for large times decays as
$t^{-\Psi}$, with $\Psi = 3/2$ for a pure system. Though its value for
a random system is not known, it is expected to be smaller than the
pure one due to the disorder induced effective attraction.  For the
polymer problem, the beads are connected and therefore this
independent particle result gives an upper bound to the decay of
overlap $q$ for the polymer problem. The data for the pure system in
Fig 2A can be fitted over the whole range by $q=.6 t^{-x}$, with
$x=1/3 <3/2$.

The aging effects we have studied might be realized experimentally
also by letting DNA equilibrate in a random medium for a certain time
and then suddenly changing the pH to start unbinding of the molecule.
Early evolution of this unbinding will shed light not only on the
dynamics of unbinding of DNA but also on the dynamics of random
systems in general.

In summary, we studied the aging effect of unbinding of a double
stranded (directed) polymer where the focus has been on the interchain
interaction. The more time the double stranded molecule spends in the
random medium the slower is the unbinding of the two strands.  We have
shown that the evolution of the overlap of the two chains has a
scaling property where the time gets scaled by the waiting time in the
random medium before unbinding.  The average number of contacts of the
two chains at early times evolve in the same manner as one of the
strands as measured by the memory of its initial configuration, with a
nonuniversal exponent that depends on the strength of the interaction.
The late time decay that reflects the true nonequilibrium behavior
shows also a power law behavior. Longer simulations are needed to
clarify this nonequilibrium dynamics.

Financial support by the Indo-German project 1L3A2B is gratefully
acknowledged. SMB also acknowledges partial support from DST
SP/S2/M17/92.

\end{multicols}
\newpage
\begin{figure}
\psfig{file=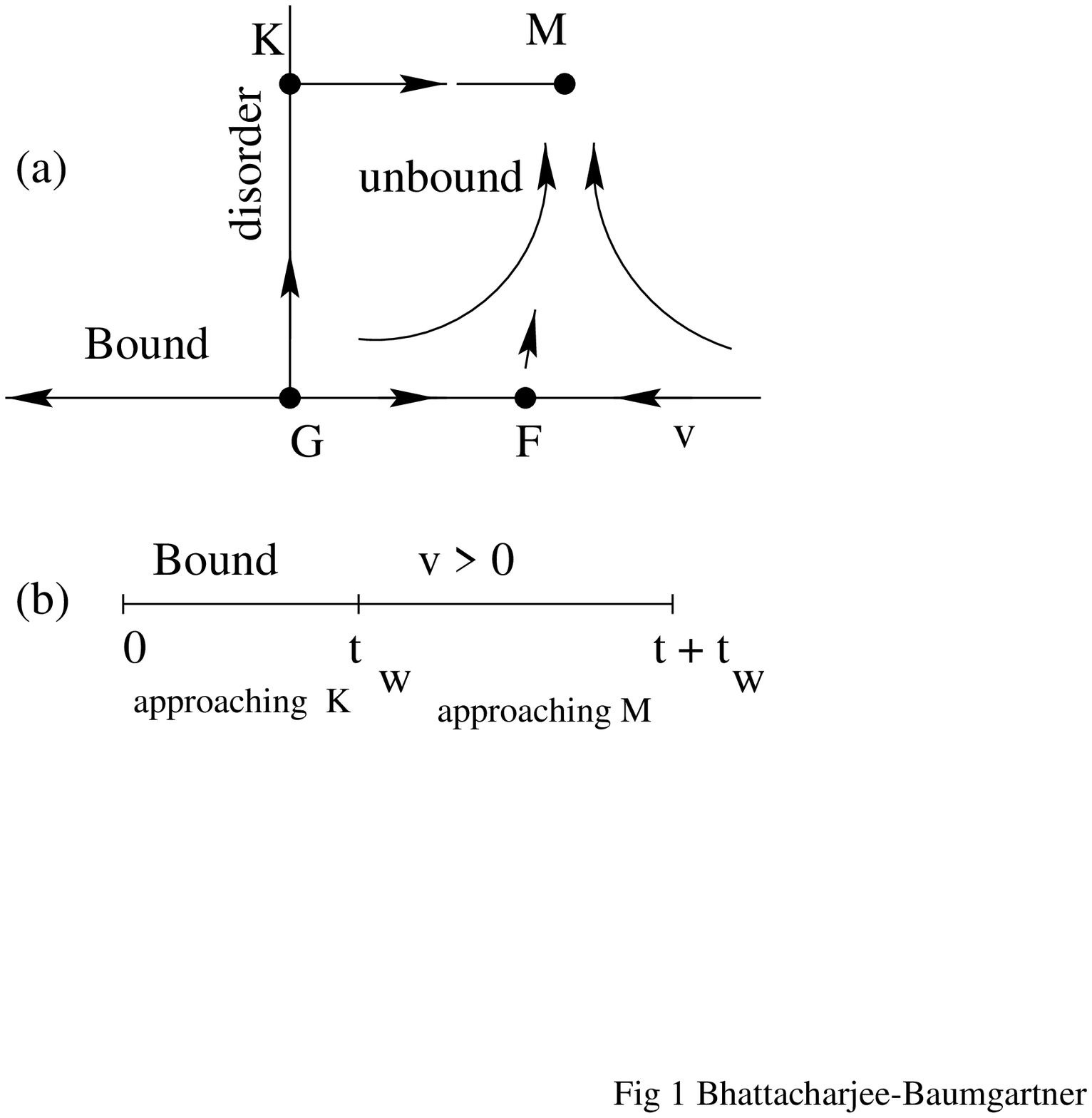}
\caption{(a)Renormalization group fixed point and flow diagram for two
chains.  G corresponds to the free pure gaussian polymers, F: pure,
repulsive (fermionic or vicious walker) polymers, K: Strong disorder
phase, M: repulsive polymers in random medium. The arrows indicate the
flows of disorder and the interaction under renormalization.(b) The
time sequence adopted in the simulation.}
\end{figure}
\newpage
\begin{figure}
{\vbox{\vskip 3cm
\psfig{file=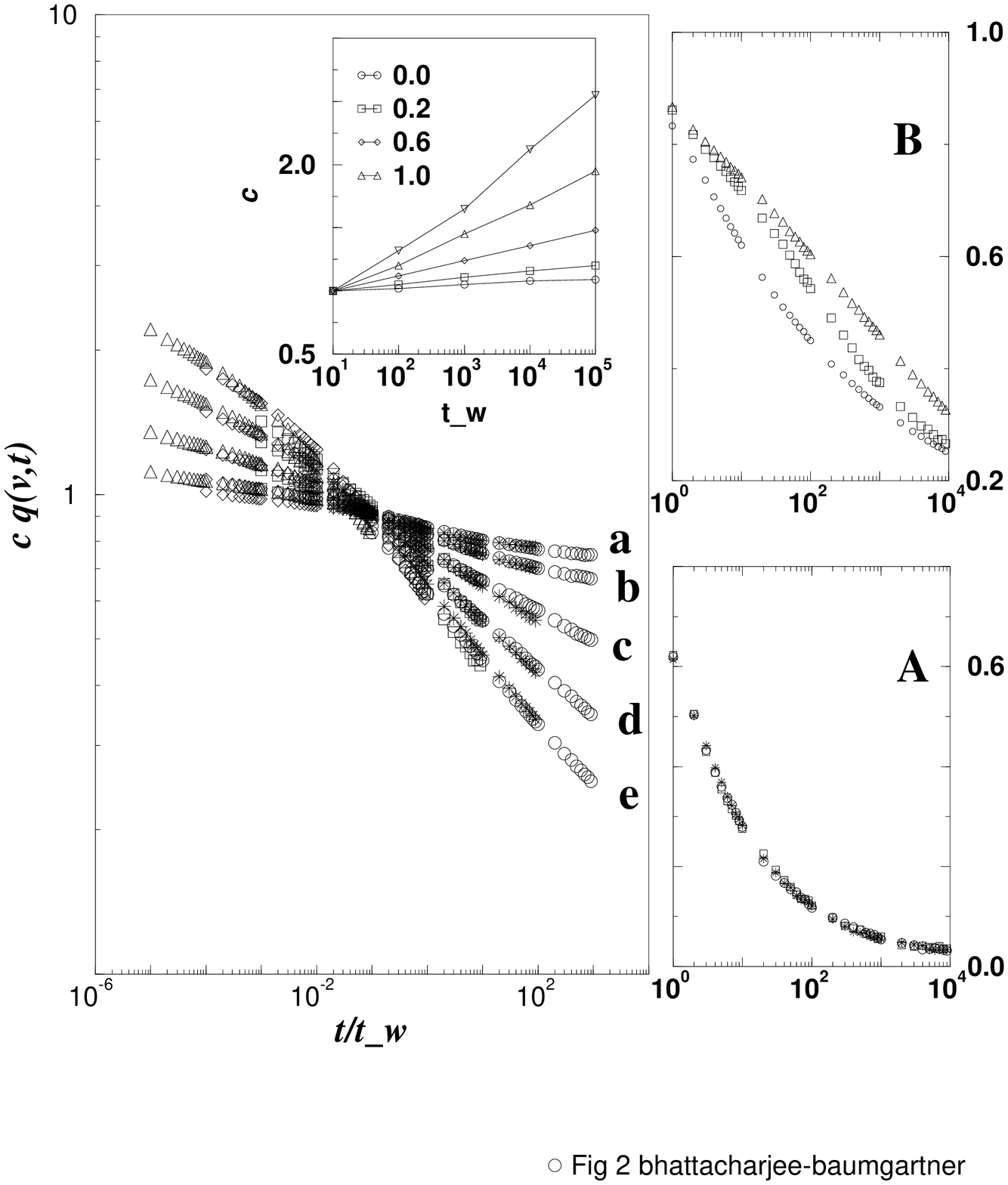,bbllx=48pt,bblly=91pt,bburx=563pt,bbury=550pt}
\caption{(A) The overlap $q(v,t+t_w)$ vs $t$ for a pure system with
$v=1.$ and $t_w = 10,\ 10^2,\ 10^3,\ 10^4$. (B) $q(v,t+t_w)$ vs $t$
for $v=1.4$ for a random system with $t_w = 10,\ 10^3, \ 10^5$. (C) $
c q(v,t)$ vs $t/t_w$ for (a) $v=0$,(b) $v=.2$, (c) $v=.6$, (d)
$v=1.0$, and (e) $v=1.4$.  The value of $c$ is chosen for each data
set (i.e., for each $v$ and $t_w$.  $t_w$ is taken as $10,10^2,10^3,
10^4$ and $10^5$. The inset shows the variation of $c$ with $t_w$ for
$v=0,\ .2,\ .6,\ 1.,\ 1.4$, $v$ increasing upwards. \label{fig:cq}}
}}
\end{figure}
\newpage
\begin{figure}
{\vbox{\vskip 3cm
\psfig{file=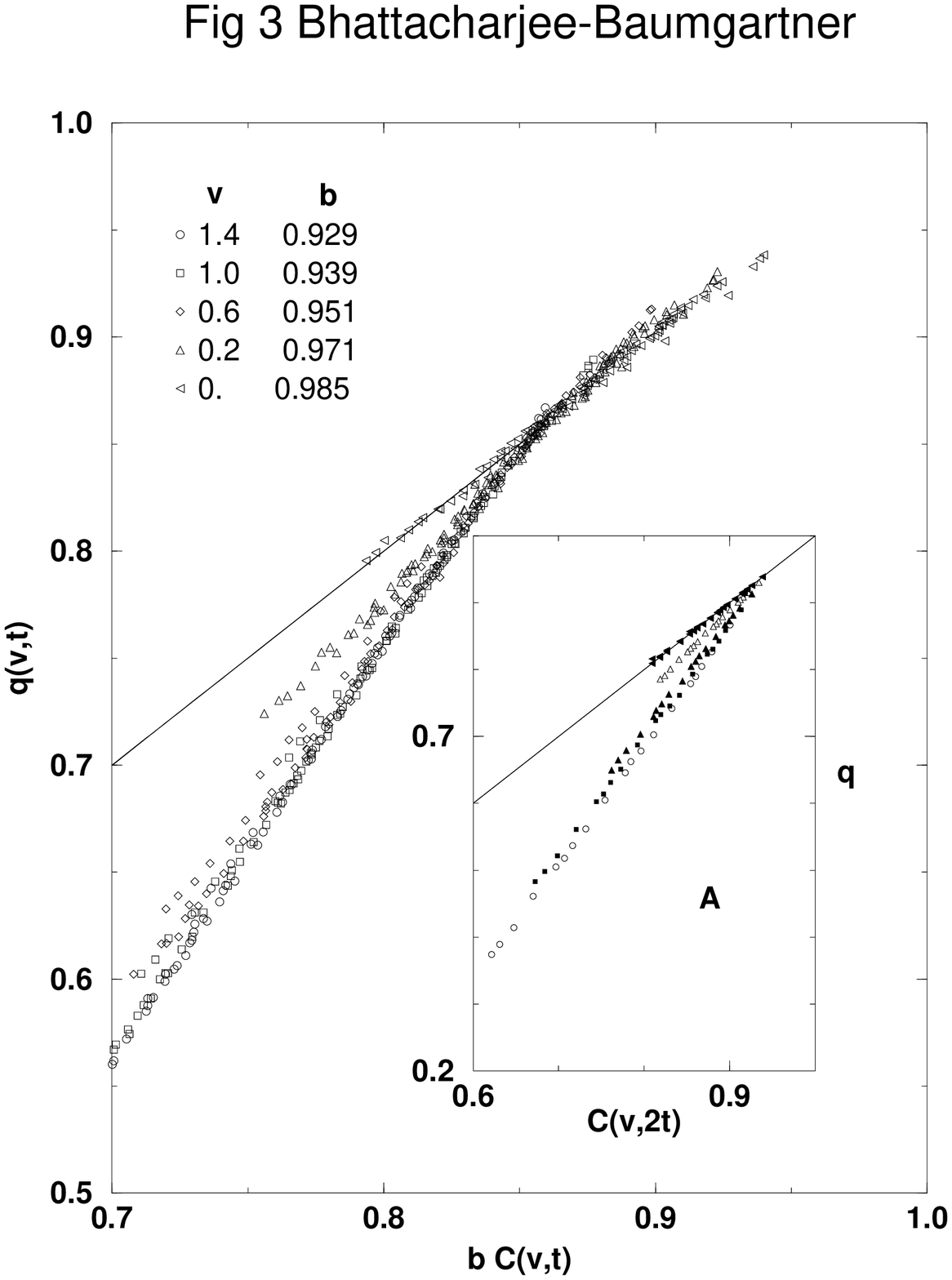,bbllx=48pt,bblly=21pt,bburx=563pt,bbury=650pt}
\caption{Plot of $q(v,t+t_w)$ vs $ b C(v,t+t_w)$ for $t < t_w$, for
various $v$ and $t_w$. Inset A shows the plot of $q(v,t+t_w)$ vs
$C(v,2 t)$, for the largest $t_w$ for each $v$ (only a few data points
are shown).  The straight line is the equality line satisfied at
$v=0$. \label{fig:qbc}}
}}
\end{figure}


\begin{references}
\bibitem{comm} The case of both ends free would correspond to annealed
disorder because the chains then explore the whole space to choose
the optimum position.
\end{references}
\end{document}